\documentclass[prd,letterpaper,onecolumn,amsmath,amsfonts,amssymb,nofootinbib]{revtex4}
\usepackage {amsmath}
\usepackage[dvips]{graphicx}
\usepackage{feynmp}
\usepackage{amssymb}
\newcommand{\be}{\begin{equation}}
\newcommand{\ee}{\end{equation}}
\newcommand{\bear}{\begin{eqnarray}}
\newcommand{\eear}{\end{eqnarray}}

\newcommand{\vev}[1]{\left\langle #1\right\rangle}

\newcommand{\lapproxeq}{\lower .7ex\hbox{$\;\stackrel{\textstyle  
<}{\sim}\;$}} 
\newcommand{\gapproxeq}{\lower .7ex\hbox{$\;\stackrel{\textstyle  
>}{\sim}\;$}} 
\newcommand{\stackdown}[2]{\lower 1.4ex\hbox{$\;\stackrel{\textstyle{#1}}  
{\scriptstyle{#2}}\;$}}
\newcommand{\beq}{\begin{equation}} 
\newcommand{\eeq}{\end{equation}} 

\newcommand{\ba}{\begin{eqnarray}}
\newcommand{\ea}{\end{eqnarray}}

\newcommand{\bea}{\begin{eqnarray}}
\newcommand{\eea}{\end{eqnarray}}

\makeatletter 
\def\slash{\@ifnextchar[{\fmsl@sh}{\fmsl@sh[0mu]}} 
\def\fmsl@sh[#1]#2{%
  \mathchoice 
    {\@fmsl@sh\displaystyle{#1}{#2}}%
    {\@fmsl@sh\textstyle{#1}{#2}}%
    {\@fmsl@sh\scriptstyle{#1}{#2}}%
    {\@fmsl@sh\scriptscriptstyle{#1}{#2}}} 
\def\@fmsl@sh#1#2#3{\m@th\ooalign{$\hfil#1\mkern#2/\hfil$\crcr$#1#3$}} 
\makeatother 
\begin{document}
\begin{flushright} 
\parbox{4.6cm}{UA-NPPS/BSM-1-11 }
\end{flushright}
\title{Dilaton dominance in the early Universe dilutes Dark Matter relic abundances }
\author{A. B.~\ Lahanas}
\email{alahanas@phys.uoa.gr}
\affiliation{University of Athens, Physics Department,  
Nuclear and Particle Physics Section,  
GR--15771  Athens, Greece}
\vspace*{2cm}
\begin{abstract}
The role of the dilaton field and its coupling to matter may result to a dilution of Dark Matter (DM ) relic densities. This is to be contrasted with quintessence scenarios in which relic densities are augmented, due to modification of the expansion rate,  since Universe is not radiation dominated at DM decoupling.  Dilaton field, besides this, affects relic densities through its coupling to dust which tends to decrease relic abundances. Thus two separate mechanisms compete each other resulting, in general, to a decrease of the relic density. This feature may be welcome and can rescue the situation if Direct Dark Matter experiments point towards small neutralino-nucleon  cross sections, implying small neutralino annihilation rates and hence large relic densities, at least in the popular supersymmetric scenarios. In the presence of a diluting mechanism  both experimental constraints can be met. The role of the dilaton for this mechanism has been studied in the context of the non-critical string theory but in this work we follow a rather general approach assuming that the dilaton  dominates only at early eras long before Big Bang Nucleosynthesis. 
\end{abstract}
\maketitle
\section{Introduction}

The nature and origin of the Dark matter (DM) of our Universe is one of the
big mysteries of modern Cosmology whose resolution is still pending. Analyzing the  
data cumulated from various observations over the past twelve years it is found that 96\% of the Universe energy budget today  consists of unknown entities, 23\% of which is Dark Matter  and 73\% Dark Energy
(DE), or vacuum  energy, which is responsible for the current acceleration of the Universe. 
These data include, among other, observations of the Universe
acceleration, using type-Ia supernovae~\cite{snIa}, measurements of cosmic microwave
background~\cite{cmb,7yrwmap} anisotropies, baryon oscillation~\cite{bao} and weak
lensing data~\cite{lensing}. 
The aforementioned results follow from best-fit analyses of various astrophysical data to the Standard Cosmological Model ($\Lambda$CDM) which can successfully describe the evolution of our Universe. 
The model is based on a Friedmann-Robertson-Walker (FRW) cosmology, involving cold
DM, at a percentage 23\% , baryonic matter at 4 \% and a positive cosmological constant
$\Lambda>0$ which is put in an ad-hoc manner in an attempt to describe the vacuum energy density.

Supersymmetry provides one of the leading DM candidates, the neutralino which is still 
lacking experimental verification. Its thermal abundance, calculated in the context 
of the simplest supersymmetry models (minimal supersymmetric model embedded
in minimal supergravity~\cite{msugra}), is severely restricted by
cosmic microwave background data. In the near future incorporating data from collider experiments, such as the LHC 
\cite{neutralino}, these models may be possibly ruled out.
However the existence of scalar fields in the primordial Universe, that contribute to the energy density, may play a dramatic role and upset the whole scenery. The quintessence \cite{quint} has been invoked in an attempt to explain the vacuum energy, in the sense that the energy it carries today is the vacuum energy measured in astrophysical observations. Its existence affects the relic abundances if Universe is not radiation dominated during DM decoupling. 
In fact DM relic density is predicted to be enhanced \cite{Salati:2002md}, and in some cases this enhancements reaches 
$ \sim 10^6 $ or so \cite{ullio}, for a review see for instance \cite{quint2}. 
In general, modifications of the expansion rate and departures from the standard cosmological scenarios may have dramatic consequences for the DM relic density \cite{Kamionkowski:1990ni,nonstand} and the observed amount of DM puts constraints on possible modifications of the Universe expansion at early eras \cite{drees}. 
A particular class is the tracking quintessence scenario  in which the   quintessence field is in a kination-dominated phase at early eras \cite{quint3}. In this context the predictions for the gravitino and axino DM are considered in \cite{pallis} while in \cite{pallis2} the predictions for the neutralino DM relic, in the popular supersymmetric schemes, is discussed in the light of  the constraints arising from the observed 
$ e^{\pm} $-spectrum by PAMELA \cite{pamela} and Fermi-LAT observations \cite{fermi}. 

In some string inspired scenarios, with a time-dependent dilaton-$\phi$ sources~\cite{elmn},
whose evolution is dictated by non-equilibrium string dynamics~\cite{cosmo},
the amount of thermal neutralino relic abundance is diluted by factor of ${\cal{O}} (10)$, 
relative to that calculated within the $\Lambda$CDM-minimal supergravity
cosmology and such models are found to survive the stringent tests of LHC~\cite{dutta}.
The dilution is due to the appearance of a friction-like term,
on the right-hand side of the appropriate Boltzmann equation. This term plays also a significant role in other considerations studied in  \cite{Mavromatos:2010nk}. 

In this paper we argue that the mechanism for the dilution of DM relic abundances is more general and can hold in other instances too, having its basis on more general features of the dilaton dynamics prevailing in early eras, independently of the non-criticality of the underlying string theory. In order to model the dilaton behavior we assume the existence of  exponential-type potentials occurring in a wide class of quintessence scenarios, supergravity models, or arise in string theories from quantum corrections. The presence of such a dilaton field, that dominates over radiation long before nucleosynthesis, affects the predictions for relic abundances in a dramatic way. In fact the conventional calculations get smaller by factors as small as  $\, {\cal{O}}(10^{-2}) \,$, in some cases, allowing therefore for smaller annihilation cross sections in the popular supersymmetric schemes employed in literature. This may alter the potential of discovering supersymmertry at collider experiments since the parameter space  allowed by the cosmological data moves to regions that would be otherwise forbidden. 
As far as other ways of discovering DM are concerned ( for reviews see \cite{dmrev}), the small cross sections required to explaine the cosmological data may affect the predictions for direct \cite{direct} and indirect 
\cite{pamela,fermi,HESS} DM searches ( for a review see \cite{conrad} ). 

\section{Setting up the model}
Omitting radiation and matter contributions the equations of motion for a time-dependent dilaton are
\begin{eqnarray}
&& \ddot \phi + 3 \, H \, \dot \phi + V^{\,'}(\phi) \;=\; 0 \nonumber \\
&& 3 \, H^2 \;=\; \dfrac{{\dot \phi}^2}{2} + V(\phi) \nonumber \\ 
&& 2 \, \dot H \;=\; - \, ( \, \varrho_\phi + p_\phi \,) \;=\; - \, {\dot \phi}^2 \label{phieq}
\end{eqnarray}
In these equations the field $\,\phi  $  is dimensionless and the potential carries dimension $mass^2$. The first of these equations is not independent but is derived from the other two. 
In order to model the dependence of the dilaton as a function of 
$\; ln \; a$, where $\; a(t) \; $  is the cosmic scale factor, we assume a linear in $\; ln \; a$ form during early eras,  which can follow from exponential-type  potentials  $\; V \sim e^{- \, k \, \phi} \; $. Such potentials are inspired by 
quintessence scenarios and they can also occur in string theories as perturbative or non-perturbative corrections. The dilaton is then given by 
\begin{eqnarray}
\phi \;=\; c \;  ln \, \left( \dfrac{a}{a_I} \right) \;+\; \phi_I \qquad ,
\label{phi}
\end{eqnarray}
where $\; c \;$  is a constant and  $\; a_I \equiv a(t_I) \;$  is the cosmic scale factor at the maximal reheating temperature reached after inflation, denoted by $\; T_I \;$, which occurred at the time $\; t_I \;$. This holds in epochs $ t < t_X $
in which dilaton dominates over radiation and matter, during DM decoupling which occured earlier than BBN. 
We know that BBN took place when $\; ln \, (a_{BBN} / a_0 )\,\simeq -22.5 \;$, corresponding to $\, T_{BBN} \simeq 1\, MeV \, $, and DM decoupling occured at a 
temperature between  $\, T_{DM} \simeq \, 5 -  20 \, GeV \,$, as dictated by interpreting DM to have supersymmetric nature, corresponding to a value of 
$\; ln \, (a_{DM} / a_0 ) \;$ between $\; \simeq -31.5  \;$ and $\; \simeq  -33.0 \;$. 
At times $ t_X $ radiation also starts contributing to the energy-matter density and at BBN must overwhelm dilaton's energy.
Therefore a reasonable region for which (\ref{phi}) holds is set by 
$\,  a \leq a_X  \,$   with $\; ln \, (a_X / a_0 )\,\simeq -25 \;$ or smaller. 
  
Beyond  $ t_X $ the dilaton is assumed to receive an almost constant value. 
The constancy of $\,\phi  $ when  hadrons are non-relativistic is rather mandatory if we do not want the diluting mechanism to 
affect the abundances of the known hadrons and especially nucleons. 
This pushes the bound on $\,  a_X  \,$, defined earlier, to even lower values.  
In fact the couplings of dilaton  to matter density is through the appearance of dissipative terms   
$\; \sim \,( \varrho_m \, - 3 \, p_m ) \, \dot{\phi}  \;$, which modifies the  continuity equation for matter,  
and such terms are vanishing when hadrons are relativistic, that is 
at temperatures higher than about $\, T_h \sim 1 \, GeV  $, coresponding to $\; ln \, (a_h / a_0 ) \,\sim -30 \;$. Below 
$\, T_h  \,  $ however hadrons are non-relativistic and dilaton couples to hadrons as $\; \sim \, \varrho_m \, \dot{\phi}  \;$. Therefore in this temperature regime the dilaton has to be almost constant in order to suppress its coupling to hadronic matter. A reasonable value is  $\, T_h = \Lambda_{QCD}  \,  $, with $ \Lambda_{QCD} \simeq 260 \, MeV \, $ the characteristic QCD scale, which pushes the bound  set on $\,  a_X  \,$ to $\; ln \, (a_X / a_0 )\,\simeq - 28.4 \;$, although   
larger values for $\, T_h  \,  $, corresponding to smaller values of $\; ln \, (a_X / a_0 ) \;$, are not excluded. 
Such values for $\, T_h  \,  $ are within the range that the coupling of the dilaton to supersymmetric matter is non-vanishing and this may have dramatic effects for the  DM relic abundances as we shall see 
{\footnote{\;
We are aware of the fact that a constant dilaton in this range cannot account for a change 
$\; \Delta \alpha / \alpha \, \sim 10^{-5}  \;$  over cosmological time scales of the fine structure constant. Interpreting 
the constancy of the dilaton as small quantum fluctuations $\, \Delta \, \phi << 1 \,$ it puts a lower limit on the couplings of dilaton to matter approaching the capability of E{\"{o}}tvos-like experiments.
}}
. 

For  $ t < t_X  $ the time derivative of $\,\phi  \;$ is related to the expansion rate by 
$\, \dot \phi = c \, H  \,$ and when this is plugged into the third of equations (\ref{phieq}) 
it can be solved for the expansion rate $\, H \,$ yielding 
\begin{eqnarray}
H^{-1} \;=\; H_I^{-1} + \dfrac{c^2}{2} \,(\, t - t_I \,) \quad .
\label{hub}
\end{eqnarray}
In this and the following equations the 
subscript I denotes quantities evaluated at $\; t_I \;$. For the expansion rate solving $\; H = \dot a / a \,  $, and using 
(\ref{hub}), we get 
\begin{eqnarray}
a \;=\; a_I \, {\left( \dfrac{c^2 \, H_I}{2} ( t - t_I ) + 1  \right)}^{2/c^2}
\label{aaa}
\end{eqnarray}
and therefore the time-dependence of the dilaton in this era is 
\begin{eqnarray}
\phi \;=\; \dfrac{2}{c} \, ln \, {\left( \dfrac{c^2 \, H_I}{2} ( t - t_I )  + 1 \right)}
+ \phi_I  \quad .
\label{phit}
\end{eqnarray}

Knowing the dilaton and the Hubble rate from the second of Eqs. (\ref{phieq}) the form of the potential can be derived
\begin{eqnarray}
V(\phi) \;=\; 
\left( \frac{6 - c^2}{2}  \right) \; H_I^{\,2} \; e^{ - \, c \, ( \phi - \phi_I ) } \quad .
\label{pot}
\end{eqnarray}
This holds in the region where the dilaton depends linearly on $\, ln \, a  \;$, as in shown in Eq. (\ref{phi}), that is 
for values of the cosmic scale factor $\, a < a_X\,$. 
If the maximum reheating temperature attained is  $\; T_I \,=\,10^{\,9}\,GeV\;$ the value of the cosmic scale factor at $\; t_I \;$ is $\; ln \, (a_I / a_0 )\,\simeq -50.86 \;$ where $\; a_0 = a(t_{today}) \;$ denotes its value today
{\footnote{\; 
We have in mind a supersymmetric model with the MSSM content whose sparticle mass spectrum is in the TeV range. Under these circumstances at the temperature $T_I$ supersymmetric as well as SM particles are all relativistic and the effective number of degrees of freedom is $\, g_{eff} = 228. 75  \,$, independently of the precise sparticle mass spectrum. This entails to the value $\; ln \, (a_I / a_0 )\,\simeq -50.86 \;$ quoted above.
}}
.
Therefore the region of applicability of Eq. (\ref{phi}) is for values of $\;a\;$ satisfying  $\; ln \, (a / a_I )\,\leq B \;$ where $\,B\,$ is set by 
$\, B = ln(a_X/a_0) + 50.86 \,  $, that is a number $\sim 20$ or so depending on the value of $a_X$. At the end point 
$a_X$, for which $\; ln \, (a_X / a_I )\,= B \;$, the potential is exponentially suppressed 
\begin{eqnarray}
V \sim \exp \,[ \,{-c \, ( \phi_X-\phi_I )}\, ] \sim \exp \,(\,{- B \,c^2 \,) }
\label{exxx}
\end{eqnarray}
independently of the sign of the constant $c$,  provided the  value of $ |c| $ is not exceedingly small. 
Since we have in mind a positive potential which drops as Universe expands the constant $c$ should be bounded by $\, c^2 \leq 6  \,$ 
as is evident from (\ref{pot}). 
As we shall see shortly, dominance of the dilaton energy over radiation is achieved for $\,c^2 > 4 \,$ and therefore the value of the potential at $a_X$, given by Eq. (\ref{exxx}), is very much suppressed  before  Nucleosynthesis, that is for times earlier than $\; t_{BBN} \;$ corresponding to $\; ln \, (a_{BBN} / a_0 )\,\simeq -22.5 \;$. 

The ratio of the kinetic to the potential energy of the dilaton field in the regime $ t<t_X $ is constant. This follows from the second of Eq. (\ref{phieq}) and the fact that 
$\, \dot{\phi} = c \, {H}  \,$, due to Eq. (\ref{phi}). In fact 
\begin{eqnarray}
V(\phi) \, / \, {\left( \frac{ {\dot{\phi}}^2}{2}  \right) } \;=\; 
 \frac{6}{c^2} \, - \,1  
\label{vover}
\end{eqnarray}
This yields a ratio of dilaton  to radiation energy density given by 
\begin{eqnarray}
\frac{  \hat{\rho}_{\,\phi}  }{  \hat{\rho}_{\,r} }
\,=\, 
\frac{m_P^2}{\hat{\rho}_{\,r}^{\,0}} \;  \frac{1}{1 - c^2/6} \; {\left( \frac{a}{a_0} \right) }^4 \; V(\phi) \quad .
\label{ratior}
\end{eqnarray}
In this we have reinstated dimensions and  hatted densities carry  dimension ${\mathrm{energy}}^{4 }$. The zero subscripts denote the corresponding quantities today. Eq. (\ref{ratior}) can be also cast in the form  
\begin{eqnarray}
\frac{  \hat{\rho}_{\,\phi}  }{  \hat{\rho}_{\,r} }
\,=\, 
\frac{3\,H_0^{\,2} \,m_P^2}{\hat{\rho}_{\,r}^{\,0}} \; \left(    \, \dfrac{H_I^{\,2}}{ H_0^{\,2}} \right)   \,     \; 
{\left( \frac{a_I}{a_0} \right)}^4 \; {\left( \frac{a}{a_I} \right)}^{\,4 - c^2 }
\label{ratior2}
\end{eqnarray}
if one expresses the potential (\ref{pot}) in terms of the cosmic scale factor $ \, a \,$. Since $\, a > a_I \,$ this ratio decreases for values of  $\, 4 < c^2 \,$ .

\begin{figure}
\begin{center}
\includegraphics[width=7cm,height=7.4cm]{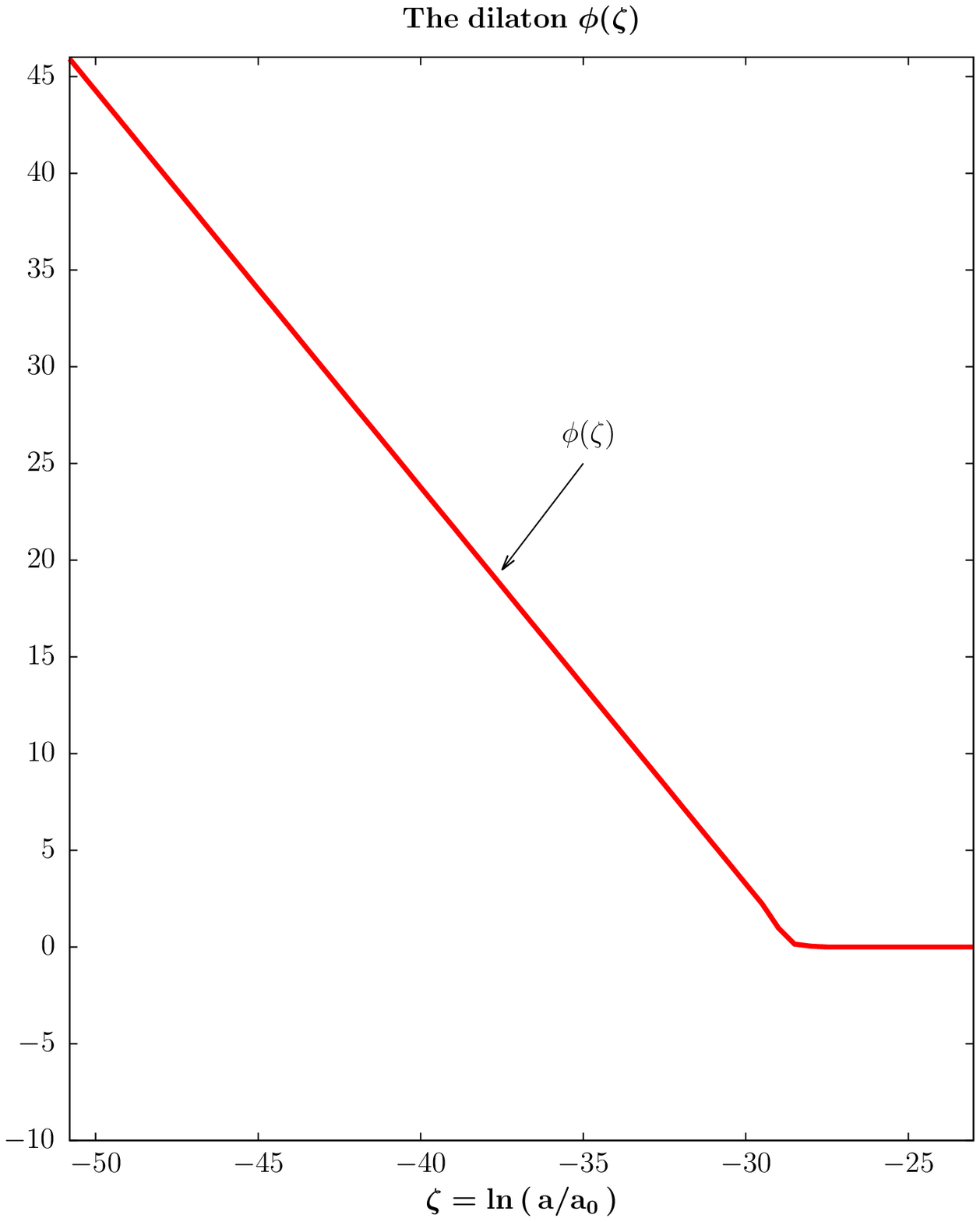}
\includegraphics[width=7.4cm,height=7.4cm]{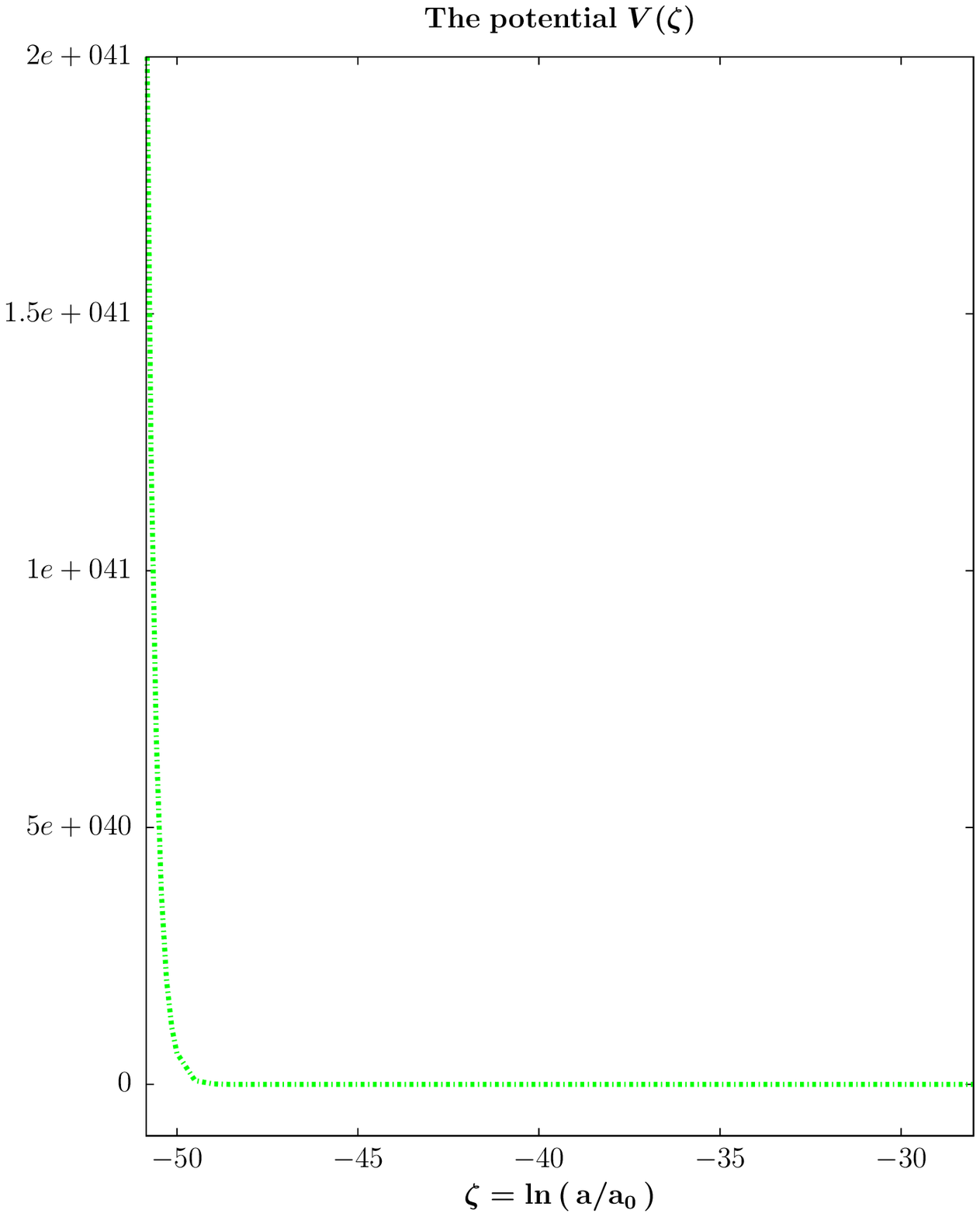}
\end{center}
\caption[]{
The dilaton (left) and the potential (right) as functions of $\; ln(\, a/a_0 \,) \;$. The ratio of dilaton's kinetic energy  to its potential energy density stay constant and thus  the fast exponential  drop-off of the dilaton potential indicates  that  the dilaton to radiation energy ratio has been suppressed before Nucleosynthesis. 
}
\label{fig0}  
\end{figure}

The behavior of the dilaton and the potential as functions of $\; ln(\, a/a_0 \,) \;  $ are shown in Fig. \ref{fig0} for a particular values of the slope $\, c \, $, in the range $ \, 4 < c^2 < 6\, $, and $\, ln( \,a_X/a_0 ) = -28.4  \,$. 
Without loss of generality the value $\phi_X $ of dilaton at the end of the dilaton-dominance period  has been taken vanishing. Actually physics results depend on the difference $ \phi - \phi_I $ so a non-vanishing value for $\phi_X $ corresponds to a different initial condition $\phi_I $ for the dilaton field. 
The ratio of the potential energy, at the end of the dilaton-dominated era, to the same energy at reheating temperature drops by at least fourty orders of magnitude. 

Since dilaton energy dominates over radiation energy in this regime the value of the Hubble rate at reheating temperature is constrained by Eq. (\ref{ratior2}). In order to quantify this suppose the dilaton to radiation energy density, at a given reheating temperature $\,T_I \,$, is 
\begin{eqnarray}
{\left( \frac{  \hat{\rho}_{\,\phi}  }{  \hat{\rho}_{\,r} } \right)} \bigg{\arrowvert}_I \, = \, 
10^{\,p}
\label{power}
\end{eqnarray}
then Eq. ( \ref{ratior2}) yields  
\begin{eqnarray}
H_I / H_0 \, = \, \left( \dfrac{0.703}{h_0} \right) \times 10^{\,42 + p/2} 
\label{hiho}
\end{eqnarray}
if $\,T_I = 10^9 \, GeV\,$. 
If inflation is responsible for the generation of the power spectrum of the curvature scalar $P_s$   and tensor $P_T$  perturbations then an upper bound on the inflationary potential, and hence on the corresponding Hubble rate at the end of inflation $H_I$, can be derived \cite{Smith:2005mm}. Assuming that tensor perturbations are small in comparison with the scalar ones the bound imposed on $H_I$ is    $\, H_I \leq \dfrac{\pi}{\sqrt{2}} \, m_P \, P_s^{1/2} \,$ which in turn yields
$\, H_I \leq 2.65 \times 10^{14} \, GeV  \,$. This  entails to the following upper bound for the ratio $H_I/H_0$,
$$
H_I / H_0 \, < \, \dfrac{1.24}{h_0} \,   \times 10^{\,56} \quad .
$$
Then on account of  (\ref{hiho}) an upper bound on $\, p \, $ is derived,
\begin{eqnarray}
p < 28.5
\label{bound}
\end{eqnarray}
This merely indicates that the ratio of the dilaton to radiation energy density (\ref{power}) can be indeed large for reasonable values of the initial conditions set at $\, T_I \,$, consistent with the bounds put on $\, H_I  $. 

From Eq. (\ref{ratior2}) the ratio $\, {  \hat{\rho}_{\,\phi}  }/{  \hat{\rho}_{\,r} } \,$ can be expressed in terms of its value at $\, T_I \,$ through 
\begin{eqnarray}
\frac{  \hat{\rho}_{\,\phi}  }{  \hat{\rho}_{\,r} }
\,=\, {\left( \frac{  \hat{\rho}_{\,\phi}  }{  \hat{\rho}_{\,r} } \right)} \bigg{\arrowvert}_I \; 
 {\left( \frac{a}{a_I} \right)}^{\,4 - c^2 }
\label{ratior3}
\end{eqnarray}
and if this ratio at $\,T_I$ is as given in Eq. (\ref{power}), the corresponding ratio at the end of the dilaton dominance period is 
\begin{eqnarray}
{\left( \frac{  \hat{\rho}_{\,\phi}  }{  \hat{\rho}_{\,r} } \right)} \bigg{\arrowvert}_X
\,=\, 10^{\,p} \; 
 {\left( \frac{a_X}{a_I} \right)}^{\,4 - c^2 } \; = \, 
 10^{\, p + N \,(4-c^2)  }  \quad .
\label{ratior4}
\end{eqnarray}
In this equation the constant $N$ is given by 
\begin{eqnarray}
N \,=\, log(e) \, ( \,b - r\, ) = 0.4344 \, ( \,b - r\, )
\end{eqnarray}
with   $ b$ and $ r$ given by $\,b=ln(a_X/a_0) \,$ and  $ \, r=ln(a_I/a_0) \, $ respectively. The first is an input while the second is determined 
by the input value of the reheating  temperature $\,T_I  \,$. For instance for $\,T_I = \sigma \, 10^9 \, GeV  \,$ the constant $r$  is given by   $\, r = -50.86 - ln \, \sigma  \,$ in a model with the content of the MSSM and mass spectrum much lighter than $T_I$. 
$\,N \,$ is a number between $ \, 8.20  \, $ and $ \, 11.16  \, $, when $\, ln \, (a_X / a_0) \,$ is taken within the range 
$ \, - 32.0 - -25.0  \, $, and the value of $ \, \sigma  $ specifying the reheating temperature is of the order or unity or so. At the end of the dilaton-dominance era the dilaton to radiation energy ratio drops from its initial value to  
\begin{eqnarray}
{\left( \frac{  \hat{\rho}_{\,\phi}  }{  \hat{\rho}_{\,r} } \right)} \bigg{\arrowvert}_X \, = \, 
10^{\,{p}^\prime}  \quad .
\label{ppower}
\end{eqnarray}
The power $ {p}^\prime $ must be smaller than   $ p $, but still large enough to guarantee that the above ratio is much larger than unity, so that the bulk of the total energy is carried by the dilaton in the regime $\, a_I < a < a_X  \,$. From 
Eqs. (\ref{ratior4}) and (\ref{ppower}) we deduce that  
\begin{eqnarray}
c^2 \,= \, 4+ \dfrac{p-p^\prime}{N} \; 
\label{cbound}
\end{eqnarray}
and therefore the  drop-off of the ratio $\, {  \hat{\rho}_{\,\phi}  }/{  \hat{\rho}_{\,r} } \,$ yields that 
$ \, c^2 \, $ is larger than 4. Combined with its upper bound $ \, c^2 < 6  \,$ discussed earlier one concludes that 
$c^2$ lies in the rather narrow range $4 < c^2 < 6  $. One can also utilize the relation 
(\ref{cbound}), in combination with the bound $\,c^2<6 \,$, to derive the following  bound on $ p-p^\prime $
$$
p-p^\prime \, < \, 2 \, N \; .
$$
Since $\, N \,$ is a number of order $\, \sim \, 10 \,$ the above upper bound set on $ p-p^\prime  $ leaves much room for values of  the  powers $ p ,  p^\prime $ to guarantee that the dilaton energy indeed overwhelms radiation energy  in the whole regime  $\,a_I < a < a_X \, $ , as is assumed in this scenario. 

\section{Dilution of DM abundances}

Concerning the calculation of relic densities, omitting the collision terms, the energy-matter density obeys the following equation 
\begin{eqnarray}
\frac{d\rho}{dt}
 +3 \hat{H} (\;{\rho} + p ) -\dfrac{\dot \phi}{\sqrt{2}  } \; ({\rho} - 3 p )=0 \; 
 \label{continuity}
\end{eqnarray}
where the last term is the coupling of the dilaton to the density
{\footnote{\;
The division by $ \sqrt{2} $ is due to the normalization 
of the dilaton whose kinetic energy appears as  $ \rho_\phi^{kin} = \dfrac{{\dot \phi}^2}{2} $, 
see the second of Eq. (\ref{phieq}).
}
.

Obviously for radiation the last term drops and hence when matter is relativistic this term is absent. This holds at a temperature $\, T >> m \,$ where $m$ is the mass of the particle under consideration. Including the collision terms into Eq. (\ref{continuity}) entails to the following equation for the number density 
\begin{eqnarray}
\frac{dn}{dt} + 3 H n + < v \sigma > (n^2 -n^2_{eq}) - \dfrac{\dot \phi}{\sqrt{2}  }\;n \;=\;0 \; .
\label{dndt}
\end{eqnarray}
This is suitable for describing the evolution of the number density during periods for which the particle is non-relativistic and pressure practically vanishes. During eras in which the particle is relativistic the last term drops, since the last term in Eq. (\ref{continuity}) drops too, and in this case (\ref{dndt}) receives the well-known form of Boltzmann equation, \cite{boltz}. 

For the number to entropy density ratio, $\, Y \,=\,n / s \,$,  
Eq. (\ref{dndt}) takes on the form 
\bear
\frac{dY}{dx}=\xi(x) \; \; m \vev{v \sigma}\; {( \;\frac{45\,G_N}{\pi} {g}_{eff}\;)}^{-1/2} \;( h+ \frac{x}{3} \frac{dh}{dx})\; ( Y^2-Y^2_{eq}) \;+\; S(x) \; \; Y  \quad .
\label{boly}
\eear
In this $x$ stands for $\, x = T / m  \,$ where $T$ is the photon gas temperature related to the radiation density $\rho_r \; $, which includes all relativistic particles at a given epoch, through
$$
\rho_r \;=\; \frac{\pi^2}{30} \; g_{eff}(T) \; T^4
$$
and $G_N $ is Newton's constant. 
In (\ref{boly}) the quantity $h$ stands for the entropic degrees of freedom related to the entropy density through 
$\, s = 2 \pi^3  \, T^3 \,h(T) \, / \, 45   \,$. 

The prefactor $\, \xi(x) \,$ appearing in Eq (\ref{boly}) is given by 
\bear
\xi(x) \;=\; {\left( 1 + \frac{\rho_m}{\rho_{\,r}} +  \frac{\rho_\phi}{8 \pi G_N \, \rho_{\,r}}    \right)}^{-1/2}
\label{xifactor}
\eear
while the source $\, S(x) \,$, in the same equation, is 
\bear
S(x) \;=\; - \frac{\; \, \phi^{\, \prime}}{\sqrt{2} \,x} \; \left( 1+ \frac{x}{3 h} \, \frac{dh}{dx}  \right)
\label{source}
\eear
In the expression for $\, \xi(x) \,$ above, 
$\, \rho_{\,r} , \rho_{\,m}  \,$ and $\,  \rho_\phi  \,$ are the radiation, matter and dilaton energy-densities respectively. Recall that we use a dilaton density having dimensions $ \, m_p^2 \, $, see Eq. \ref{phieq}. Note that no cosmological-term contributes to Eq. (\ref{xifactor}) since such a term is absent at DM decoupling and long after it. 
In conventional treatments the prefactor $\, \xi(x) \,$ is unity since the DM freeze-out is assumed to take place in the  radiation dominated era. However in the presence of the dilaton energy term this is smaller than unity and the density 
$\,Y \,$ decreases slower, as temperature drops, than in the conventional cases where $\, \xi(x) = 1\,$ . In the source term $\, S(x) \,$, given by Eq. (\ref{source}),  the quantity $\phi^{\, \prime} $ is the derivative of the dilaton with respect $\, ln \,( \, a / a_0) \,$ and if this is negative then the source acts in the opposite direction than $\, \xi(x) \,$ tending to decrease the density $\,Y \,$ faster as $x$ decreases. 

\begin{figure}
\begin{center}
\includegraphics[width=9.5cm]{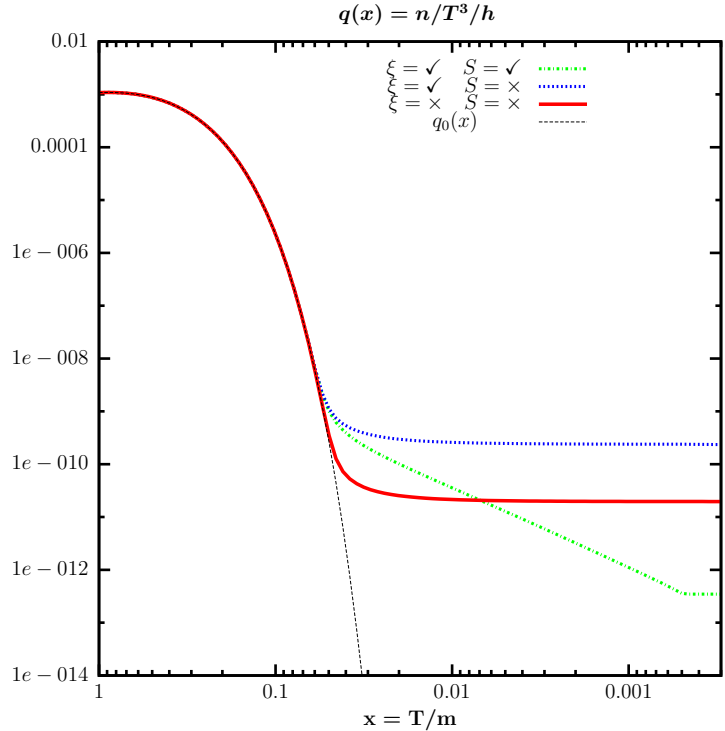}
\end{center}
\caption[]{
The LSP Dark Matter number density to entropy density ratio $ \;{q = \dfrac{n}{T^3 h} }\; $ as function of 
$\, x = \frac{T }{m_{LSP}} \, $ in a particular supergavity model. The values of $ \xi , S $ denote the status of the $\xi$-factor and the source respectively  ( $\checkmark $ for open, $\times $ for switched-off ). For comparison the corresponding equilibrium density $ q_0(x) $ has been also drawn. 
}
\label{fig1}  
\end{figure}
\begin{figure}
\begin{center}
\includegraphics[width=9.5cm]{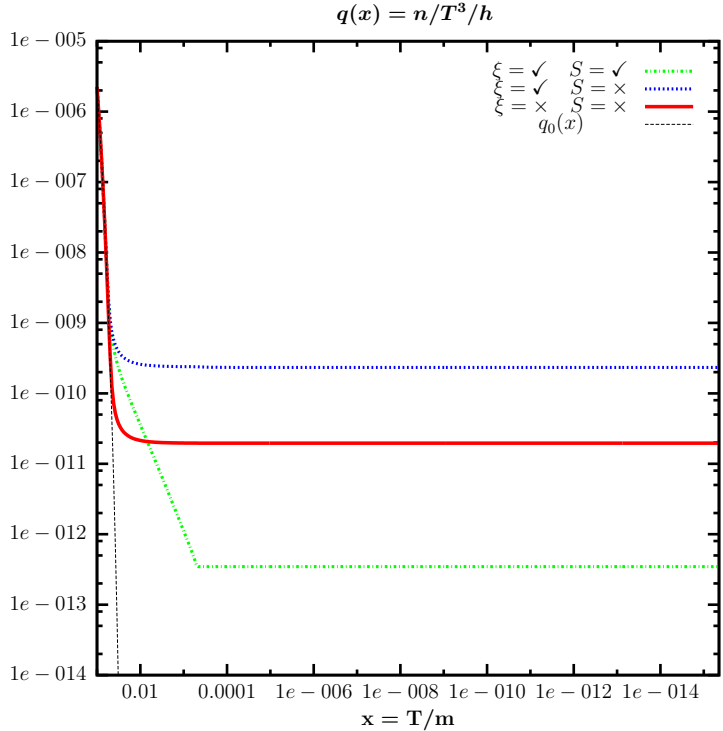}
\end{center}
\caption[]{
The same as in figure \ref{fig1} with $\, x = \frac{T }{m_{LSP}} \, $ from $0.1$ to values corresponding to $CMB$ temperature today.
}
\label{fig2}  
\end{figure}

It should be noted that for simplicity we have assumed the lowest order, in $\alpha^\prime$, contributions to the form factors 
$\; e^{-\psi(\phi)} \;$ and $\; Z(\phi) \;$ associated with the scalar curvature $R$ and dilaton kinetic terms of the effective action, in the string frame, and hence the simple expressions for the $\dot{\phi} $-dependent terms of 
Eqs (\ref{continuity}) and (\ref{dndt}). Also the dilatonic charge has been assumed vanishing. However the couplings of the dilaton to matter may evolve in time with the dilaton itself and depend on the particle species in a non-universal way. Therefore other options are available which in the string theory arise from loop corrections or non-perturbative string effects \cite{Gasperini:2001mr,Gasperini:2001pc}. In such cases the coupling of matter to $\dot{\phi} $ in the continuity equation Eq. (\ref{continuity}), which is mainly controlled by $\; \psi^\prime (\phi) \equiv d\psi/d \phi $, is not a constant. Besides there is an additional contribution to the continuity equation which depends on the dilatonic charge, if the latter is assumed non-vanishing \cite{Gasperini:2001mr,Gasperini:2001pc,Gasperini:2002bn}.  Including these effects will give rise to modified 
$\dot{\phi} $-dependent terms in Eqs. (\ref{continuity}) and (\ref{dndt}) resulting to a source  $\;S(x)$ in 
Eq. (\ref{source}) which is multiplied by $\;  \sim \psi^\prime (\phi)  $, provided the dilatonic charge of Dark Matter is 
taken vanishing. This will still tend to decrease the density $Y$, as $x$ decreases, if $\; \psi^\prime (\phi) > 0 $ in the regime following Dark Matter decoupling. In particular if $\; \psi^\prime (\phi) > 1 $ the dilution caused to the relic density is enhanced, in comparison with that caused by the source term as it appears in Eq. (\ref{source}), or gets smaller if 
$\; \psi^\prime (\phi) < 1 $. Certainly in order to further study the effects of this term one needs a better understanding of how to handle the corrections to $\, \psi(\phi) $  arising from  the underlying string dynamics. For definiteness in this work, and in order to quantify  the effect of the dilution of the abundance of  Dark Matter, we assume a gravi-dilaton effective action in the  lowest order in the string slope $\alpha^\prime$. 

The effects of the presence of the factors $ \, \xi(x)$ and $\, S(x) $, as given by Eqs (\ref{xifactor} ) and 
(\ref{source} ),  is shown in figures \ref{fig1} and \ref{fig2} where we plot the density 
as function of the temperature, actually $ x = T / m_{LSP} $, for particular SUSY inputs. Eq (\ref{boly}) has been integrated  numerically which yields more accurate results than the approximate solutions employed  in \cite{elmn}. 
The displayed figures correspond to a supergravity model with inputs given by $\, m_0=1100.0 \, GeV ,\, M_{1/2} =1200.0 \, GeV  \,$ and $\, A_0 = 0 \; GeV  \,$. We have taken $tan \beta = 40 $ and the parameter $\mu$ is taken positive, $\mu > 0 $. The value of  $\,b \equiv ln(a_X/a_0) \,$, setting the onset of the epoch after which the dilaton is constant ( $\phi = 0  $~), has been taken  $ - 28.4 $ corresponding to a temperature $\,\Lambda_{QCD} = 260 \, GeV$ as we have already discussed. For the particular SUSY inputs the  LSP Bino has a mass $m_{LSP} = 527.2 \, GeV  $ and the point $b = - 28.4$ corresponds on the $x$ axis to a value  $x \simeq 0.0005$. For comparison, except the ordinary case scenario, where both the source $S$ and the 
$\xi - $ factor are absent ( red solid line ), the cases where both $\xi $  and $S$ are open ( green dashed-dotted line ), or when only the $\xi - $ factor is present ( blue short-dashed line ) are also shown. The very thin dashed line, that rapidly drops, is the corresponding equilibrium density. In the most interesting case, that both terms are switched on, the density is monotonically decreasing after decoupling due to the appearance of the source term. The rapid change around 
$ x \simeq 0.0005 $, corresponding to $b = - 28.4 $ where dilaton reaches its  constancy, are shown in figures \ref{fig1} and \ref{fig2}. In the specific example shown the relic density is diluted by a factor of $\sim 50  $, as can be seen by comparing today's density values for the conventional case ( red solid line ) and the case where both $\xi - $ factor and the source term are present ( green dashed line ). In the first case the relic density predicted is $ \, \Omega_{LSP} \, h_0^2 \, = \, 6.059 \, $ while in the second case the relic density is considerably reduced falling in the WMAP allowed range $ \, \Omega_{LSP} \, h_0^2 \, = \, 0.1116 \, $. In general, for given $\,b = ln(a_X/a_0) \,$, one can obtain reduction factors in the range ${\cal{O}} (5 \, - \,50) \,$, the smaller (larger) corresponding to lighter (heavier) neutralino masses.  

\section{Conclusions}
In this letter we have shown that the dilaton dynamics during early eras, long before Nucleosynthesis, in conjunction with its coupling to Dark Matter may have dramatic consequences for the predicted Dark Matter relic density. Modeling the dilaton evolution to be that dictated by exponential type  potentials  $\; V \sim e^{- \, k \, \phi} \; $, occuring in quintessence scenarios and string theory,  the ordinary predicted DM density may be diluted by large factors ranging from  $\, {\cal{O}} (5) \,$ to $\, {\cal{O}} (50) \,$. This dilution mechanism is consistent with the absence of dilaton couplings to ordinary matter ( hadrons ), in the continuity equations, but it affects DM relics since dilaton dominates over radiation during and after  DM decoupling. 
This allows for LSP annihilation cross sections, in the popular supersymmetric schemes, that are smaller by an order of magnitude or more. 
This however may imply smaller inelastic cross sections of the neutralino LSP with nucleons putting farther the potential of discovering supersymmetric DM at proposed direct detection experiments \cite{direct}. 
As far as indirect detection experiments are concerned, indirect searches of Dark Matter through antimatter production has stirred much interest the last three years. The PAMELA data \cite{pamela} in combination with those provided by Fermi-LAT \cite{fermi} and HESS \cite{HESS} may be conditionally explained as DM annihilations in the galactic halo that generates the produced antiparticle flux \cite{Bergstrom:2009fa}. In the case under consideration the smaller annihilation cross sections,  required to satisfy WMAP data, makes even harder the possibility that antimatter fluxes observed in the cosmic ray are relevant to annihilation of the neutralino LSP in the galactic halo. In conventional supersymmetric models the situation can be rescued at the cost of considering large boost factors. However even in this case DM annihilation looks a rather remote explanation for interpreting the aforementioned data and other more conservative explanations exist, as for instance antimatter produced by pulsars or supernovae etc \cite{Bergstrom:2010gh}. 

A complete phenomenological study of supersymmetric models addressing all these issues is in progress and the results will appear in a forthcoming publication \cite{spanlah}.

\section*{Acknowledgments}
I wish to thank Dr. V. C. Spanos for his comments and enlightening discussions. 
This work is supported by the European Union under the Marie Curie Initial Training Network  "UNILHC" PITN-GA-2009-237920 and RTN european Programme 
MRTN-ct-2006-035505 HEPTOOLS. The author acknoweledges also support by the NKUA Special Research account.

\end{document}